\documentclass[11pt,twoside]{article}
\usepackage{asp2010}
\resetcounters
\begin{document}
\title{The MUSE Data Reduction Pipeline:\\Status after Preliminary Acceptance Europe}
\author{Peter~M.~Weilbacher$^1$, Ole~Streicher$^1$, Tanya~Urrutia$^1$,
        Arlette P\'econtal-Rousset$^2$, Aur\'elien Jarno$^2$, Roland Bacon$^2$
\affil{$^1$Leibniz-Institut f\"ur Astrophysik Potsdam (AIP), An der Sternwarte 16, D-14482 Potsdam, Germany}
\affil{$^2$Universit\'e de Lyon, Lyon, F-69003, France;
           Universit\'e Lyon 1, Observatoire de Lyon, 9 avenue Charles Andr\'e, Saint-Genis Laval, F-69230, France;
           CNRS, UMR 5574, Centre de Recherche Astrophysique de Lyon;
           Ecole Normale Sup\'erieure de Lyon, Lyon, F-69007, France}}

\begin{abstract}
MUSE, a giant integral field spectrograph, is about to become the newest
facility instrument at the VLT. It will see first light in February 2014. Here,
we summarize the properties of the instrument as built and outline
functionality of the data reduction system, that transforms the raw data that
gets recorded separately in 24 IFUs by 4k CCDs, into a fully calibrated,
scientifically usable data cube.
We then describe recent work regarding geometrical calibration of the
instrument and testing of the processing pipeline, before concluding with
results of the Preliminary Acceptance in Europe and an outlook to the
on-sky commissioning.
\end{abstract}

\section{The instrument}
MUSE \citep{BAA+12} is a wide-field integral field spectrograph built for the
ESO Very Large Telescope. It passed its Preliminary Acceptance in Europe (PAE)
on Sept.~10th, 2013, and is currently being shipped to Chile for installation
at the telescope.  MUSE consists of 24 image slicer integral field units (IFUs)
in as many subsections of the field of view. Spatial and spectral sampling of
0\farcs2 per spaxel and 1.25\,\AA\ per pixel over a field of
$\sim1\arcmin\times1\arcmin$ coupled with a spectral resolution between 1600
and 3500 ensure the usefulness of MUSE for a large variety of scientific
observations. For the study of the early universe the long wavelength range of
480 to 930\,nm allows the detection of Lyman-$\alpha$ emission in the redshift
range $3 \lesssim z \lesssim 6.6$. A high throughput of $\gtrsim$35\% with a
peak of $\sim$55\% at 7000\,\AA\ in conjunction with high stability will enable
users to carry out the deepest spectroscopic observations of the universe. In
the future, MUSE will be coupled with the adaptive optics module GALACSI
\citep{SLA+12}, and reach even fainter limits with ground-layer correction. A
narrow-field mode will use laser tomography for high-order correction and
deliver significant strehl ratios in the red part of the spectrum over a field
of $7\farcs5\times7\farcs5$.

\section{Data processing}
The broad range of applications means that the task of the MUSE {\em data
reduction system} \citep[DRS,][]{WSU+12} or {\em pipeline} is the
transformation from the raw, CCD-based data into a datacube that is corrected
for instrumental effects. For this, it uses the ESO Common Pipeline Library
\citep[CPL,][\url{http://www.eso.org/sci/software/cpl}]{MBB+04} and the typical
ESO plugin framework ("{\it recipes}").  Alternatively, the MUSE pipeline can
be run using a Python-CPL interface \citep{SW12} that can also be used with
other ESO pipelines. The consortium uses this interface to integrate the
pipeline into the Astro-WISE system \citep{PSV12}.

\subsection{Pipeline layout}
The MUSE DRS consists of two layers:
{\it 1.} Basic processing, including instrument-internal calibrations like
master bias creation and flat-fielding.  This was originally planned to be run
only using parallel processes, but thanks to updates in the CPL it recently
became possible to optionally use full internal parallelization. At the end of
this process the data has been converted from CCD-based images to pixel tables.
{\it 2.} Post-processing, including on-sky calibration exposures, e.\,g.\ flux
calibration and sky subtraction. This is internally parallelized using OpenMP
(\url{http://www.openmp.org/}) as far as possible, and converts the science
data from pixel tables into the final datacube.

The most important algorithms of the MUSE pipeline have been presented
previously, such as the single-resampling paradigm \citep{WGR+09} and a
sky-subtraction method adapted to this approach \citep{SWB+11}.

\subsection{Geometrical calibration}
\begin{figure}
\begin{centering}
\includegraphics[width=0.9\textwidth]{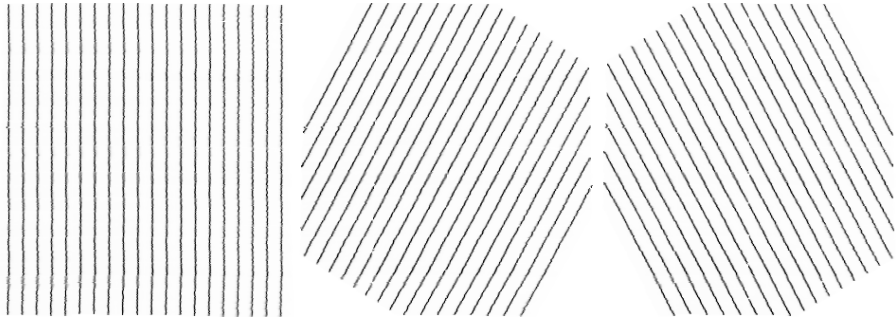}
\caption{Three slit-mask exposures used to visually verify the geometrical
         calibration. The exposures were taken with different tilts of the
         mask to cover most of the field of view. Small imperfections are
         still visible.}\label{pmw:slitmask}
\end{centering}
\end{figure}
One component that has been developed within the last year and that is now part
of the basic instrument calibration is the derivation of the {\it geometry} of
the instrument, i.\,e.\ how the 1152 slices of MUSE are distributed within the
field of view. For this, a special calibration sequence, using a multi-pinhole
mask is run, vertically shifting the mask in the focal plane so that one can
observe how the pinholes subsequently illuminate all slices.  Together with
known basic properties of the mask, the offsets of the shifts, and the
instrument design, this data allows to determine relative location (x and y
position) and angle of each slice of each IFU of MUSE. Data taken with another
mask (see Fig.~\ref{pmw:slitmask}) after the exposure
sequence can be used to verify the output of the procedure.

\section{Testing}
\begin{figure}
\begin{centering}
\includegraphics[width=0.9\textwidth]{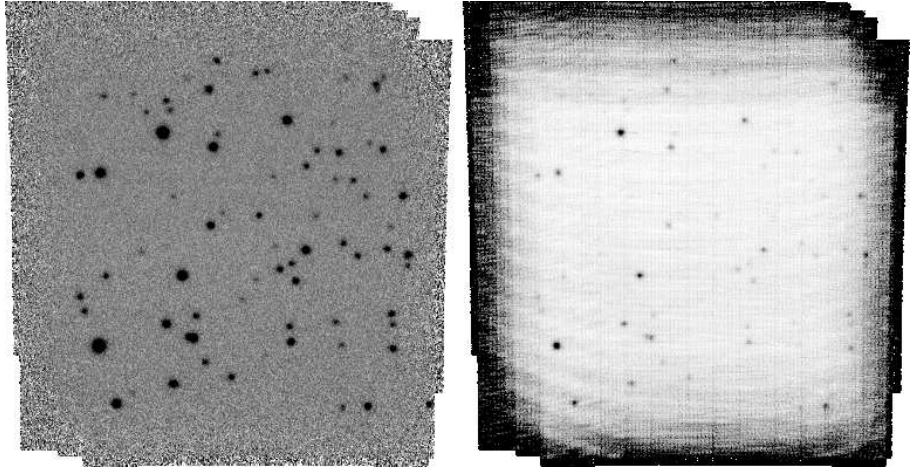}
\caption{Twenty dithered exposures of a simulated stellar field were
         successfully combined into one datacube. Here we show the cube
         plane at 6050\,\AA\ of the data part of the cube (left) and the
         corresponding variance (right). Both panels are displayed in negative
         linear greyscale. The dither pattern and some instrumental features
         are visible in the variance but not in the data.}\label{pmw:starfield}
\end{centering}
\end{figure}

The main effort in the past year was related to testing the various parts of
the MUSE pipeline. While in previous years we were able to check performance
only on single-IFU data, the instrument was completely assembled over the last
months \citep{CAA+12}, so that we could now get multi-IFU data from exposures
taken with the internal calibration unit \citep{KBH+12}. All important pipeline
calibrations (bias, flat-fields, wavelength calibration) could be verified
against specifications, and a datacube containing data from all 24 IFUs could
finally be created in June 2013.

At the same time, data of different astrophyical scenes, simulated with the
instrument numerical model \citep[INM;][]{JBP+12}, was processed by the
pipeline. The datacubes created by this were then used to support development
of higher-level analysis tools for the different science objectives
\citep{RBW+12}, such as extracting stellar spectra from dense stellar fields
\citep{KWR13} or detection of Lyman-$\alpha$ emitters (Herenz et al.~in prep.).
Such a reduced exposure, combined from 20 single exposures, is presented in
Fig.~\ref{pmw:starfield}. For more details on performance with simulated
science data see \citet{WSU+12}.

Apart from such manual testing, two automated testing methods were implemented
during the last years: A suite of unit-testing programs and an automated
nightly reduction. The former was partly developed together with the MUSE
pipeline library functions, with dedicated test data. It is automatically run
on several of the development machines, and manually before committing more
complex code changes to the development repository. It immediately reports
unforeseen problems with code changes, and code coverage by the unit tests.
The nightly reduction also checks the outer layer of the data reduction, the
{\it recipes}. It does a full data reduction (for speed reasons on a subset of
the 24 IFUs) and reports unforeseen problems. Before doing major changes, the
results of such a reduction session are saved as reference dataset. This
approach has so far successfully alerted the developers whenever code changes
had unexpected side effects, allowing for quick bug fixing (or back-outs).

A test report was delivered to ESO for PAE, which included results of the above
testing, plus a few tests designed to verify performance against the
requirements.

\section{Conclusions and Outlook}
Since the MUSE instrument, including the data reduction system, passed the PAE
and is being shipped to Paranal, attention on the DRS now turns to work
necessary to prepare the pipeline for on-sky commissioning. The geometrical
calibration described above still has to be tuned to improve its accuracy
before the instrument is pointed to the sky. Since the INM is not a perfect
representation of instrument and sky, specific points can only be addressed
with on-sky data. These include measuring the throughput curve from standard
stars, and in which way it is going to be applied to MUSE data (smoothing,
rejection of outliers). The correction for differential atmospheric refraction
has to be checked on the real sky as well. A calibration of the relative
strengths of the night-sky emission lines is necessary for an effective sky
subtraction.

\acknowledgements
PMW, OS, and TU are grateful for support by German Verbundforschung through the
"MUSE/D3Dnet" project (grant 05A11BA2).

\bibliography{P023}

\end{document}